\documentclass{usiinftr}

\usepackage{times}
\usepackage{fullpage}

\usepackage{xcolor}
\definecolor{pennblue}{cmyk}{1,.65,0,.3}
\definecolor{pennred}{cmyk}{0,1,.65,.34}

\usepackage{graphicx}
\graphicspath{{./figures/}{../figures/}}

\usepackage{amsmath}

\newcommand{\smartparagraph}[1]{\noindent{\bf #1}\ }

\usepackage{cmtt}

\usepackage{alltt}

\usepackage{paralist}
\usepackage{lastpage}
\usepackage{times}
\usepackage{tikz}
\usepackage{alltt}
\usepackage{graphicx}
\usepackage{balance}
\usepackage[font=small,format=plain,labelfont=bf,textfont=bf]{caption}
\usepackage{subcaption}
\usepackage{xspace}

\usepackage[linesnumbered,ruled,vlined]{algorithm2e}

\usepackage{url}

\usepackage{listings}

\begin{document}

\title{\bf Network Hardware-Accelerated Consensus}

\author{Huynh Tu Dang}{1}
\author{Pietro Bressana}{1}
\author{Han Wang}{2}
\author{Ki Suh Lee}{2}
\author{\\Hakim Weatherspoon}{2}
\author{Marco Canini}{3}
\author{Fernando Pedone}{1}
\author{Robert Soul\'{e}}{1}

\affiliation{1}{\USIINF}
\affiliation{2}{Cornell University, United States of America}
\affiliation{3}{Universit\'{e} catholique de Louvain,  Belgium}

\TRnumber{2016-03}

\date{}
\maketitle
\thispagestyle{empty}

\begin{abstract}
Consensus protocols are the foundation for building many
fault-tolerant distributed systems and services. This paper posits
that there are significant performance benefits to be gained by
offering consensus as a network service (CAANS). CAANS leverages
recent advances in commodity networking hardware design and
programmability to implement consensus protocol logic in network
devices. CAANS provides a complete Paxos protocol, is a drop-in
replacement for software-based implementations of Paxos, makes no
restrictions on network topologies, and is implemented in a
higher-level, data-plane programming language, allowing for
portability across a range of target devices.  At the same time, CAANS
significantly increases throughput and reduces latency for consensus
operations. Consensus logic executing in hardware can transmit
consensus messages at line speed, with latency only slightly
higher than simply forwarding packets.
\end{abstract}

\section{Introduction}

Consensus is a fundamental problem for distributed systems.  It
consists of getting a group of participants to reliably agree on some
value used for computation (e.g., the next valid application state).
Several protocols have been proposed to solve the consensus
problem~\cite{lamport98,OL88,ongaro14,CT96}, and these protocols are
the foundation for building fault-tolerant systems, including the core
infrastructure of data centers.  For example, consensus protocols are
the basis for state machine replication~\cite{lamport78,schneider90},
which is used to implement key services, such as
OpenReplica~\cite{openreplica}, Ceph~\cite{ceph}, and
Chubby~\cite{burrows06}.  Moreover, several important distributed
 problems can be reduced to consensus, such as atomic
broadcast~\cite{CT96} and atomic commit~\cite{gray06}.

Given the key role that consensus protocols play in reliable
computing, it is unsurprising that there is significant interest in
optimizing their performance.  Prior efforts towards this goal span a
range of approaches, including exploiting application semantics (e.g.,
EPaxos~\cite{moraru13}, Generalized Paxos~\cite{Lamport04}, Generic
Broadcast~\cite{PS99}), strengthening assumptions about the network
(e.g., FastPaxos~\cite{lamport06}, Speculative Paxos~\cite{ports15}),
restricting the protocol (e.g., Zookeeper atomic
broadcast~\cite{reed08}), or careful engineering (e.g.,
Gaios~\cite{bolosky11}).

This paper presents CAANS, an alternative approach to increasing the
performance of consensus protocols. Rather than implementing consensus
logic in software running on servers, CAANS allows the network to
offer consensus as a service, just as point-to-point communication is
provided today. The CAANS approach provides two specific
benefits. First, since the logic traditionally performed at servers is
executed directly in the network, consensus messages travel fewer
hops, resulting in decreased latency.  Second, rather than executing
server logic in software, which adds overhead for passing packets up
the protocol stack and involves expensive message broadcast
operations, the same operations are realized ``on the wire'' by
specialized hardware, improving both latency and throughput.

CAANS is possible because of recent trends in commodity networking
hardware. Switches and NICs are becoming more powerful, and several
devices are on the horizon that offer flexible hardware with
customizable packet processing pipelines, including Protocol
Independent Switch Architecture (PISA) chips from Barefoot
networks~\cite{bosshart13}, FlexPipe from Intel~\cite{jose15},
NFP-6xxx from Netronome~\cite{netronome-hotchips}, and Xpliant from
Cavium~\cite{xpliant}.

At the same time, importantly, these devices are becoming easier to
program.  Several vendors and consortia have developed high-level
programming languages that are specifically designed to support
network customization. Notable language examples include Huawei's
POF (protocol-oblivious forwarding)~\cite{song13}, Xilinx's PX~\cite{brebner14}, and the P4
Consortium's P4 (programming protocol-independent packet processors)~\cite{bosshart14}. Such languages significantly lower
the barrier for implementing new data\-plane functionality and network
protocols. Moreover, they offer the possibility for co-designing the
network and consensus protocols to optimize performance, as suggested
by NetPaxos~\cite{dang15}.

Recently, Istv\'{a}n et al.~\cite{istvan16} implemented Zookeeper
atomic broadcast in a Xilinx VC709 FPGA.  Their work clearly
demonstrates the performance benefits achievable by implementing
consensus logic in hardware. However, their system suffers from
several significant limitations: $(i)$ The FPGA implementation does
not provide an application-level API.  Thus, any application that
wishes to use consensus must also be implemented inside of the FPGA.
Developing an application more complex than a key-value store (such as
Zookeeper itself) might prove a daunting task. $(ii)$ The
implementation is platform-specific, and is not portable to other
network devices, such as programmable ASICs, Network Processor Units
(NPUs), or other FPGA boards. $(iii)$ Zookeeper atomic broadcast is a
restricted version of Paxos that has not been used in storage systems
other than Zookeeper.

In contrast, CAANS implements a complete Paxos protocol and is
a \emph{drop-in} replacement for software-based implementations.
Moreover, CAANS is implemented in a higher-level, data-plane
programming language, P4~\cite{bosshart14}, allowing for portability
across a range of target devices. 
 
We have evaluated CAANS on a variety of target architectures,
including three different FPGAs, an NPU, and in simulation.  Because
CAANS offers the same API as a software-based Paxos, we were able to
directly compare their performance. Our evaluation shows that CAANS
significantly increases throughput and reduces latency for consensus
operations. Consensus logic executing in hardware can transmit
consensus messages at line speed, with latency only slightly higher
than simply forwarding packets. Overall, this paper makes the
following contributions:

\begin{itemize}
\itemsep -0.5em 
 \item It describes the design and implementation of a system offering
 consensus as a network service that is a drop-in replacement for
 software-based consensus libraries.
 \item It provides a complex use-case for data-plane programming
 languages, and evaluates their performance across a range of target
 devices.
 \item It presents experiments demonstrating the performance
 improvements gained by moving consensus logic into the network.
\end{itemize}     

The rest of this paper is organized as follows.  We first provide
 short summaries of the Paxos protocol (\S\ref{sec:background}). We
 then discuss the design (\S\ref{sec:design}) and
 implementation (\S\ref{sec:implementation}) of CAANS. Next, we
 provide a thorough evaluation (\S\ref{sec:evaluation}) of our
 prototype. Finally, we discuss related work (\S\ref{sec:related}),
 and conclude (\S\ref{sec:conclusion}).

\section{Background and Motivation}
\label{sec:background}

Before detailing the design of CAANS, we briefly describe the Paxos
protocol, and present experiments that motivate moving
consensus logic into the network.

\subsection{The Paxos Consensus Protocol}

\begin{figure}[t]
\centering
 \includegraphics[width=0.4\columnwidth]{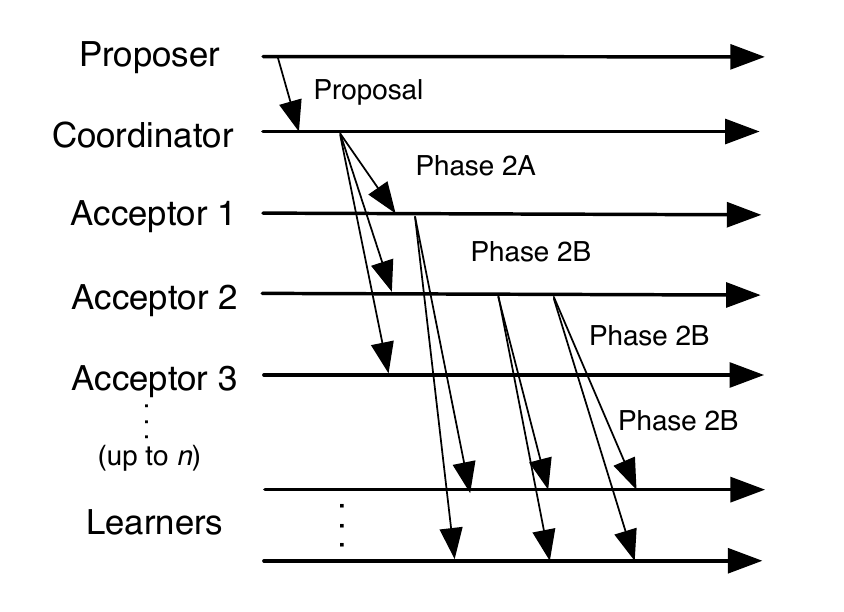}
\caption{The Paxos Phase 2 communication pattern.}
\label{fig:paxos}
\end{figure}

Paxos is a fault-tolerant consensus protocol with important
characteristics: it has been proven safe under asynchronous
assumptions (i.e., when there are no timing bounds on
message propagation and process execution), live under weak
synchronous assumptions, and resilience-optimum~\cite{lamport98}.

Paxos distinguishes the following roles that a process can play: {\em
proposers}, {\em acceptors} and {\em learners}.  Clients of a
replicated service are typically proposers, and propose commands that
need to be ordered by Paxos before they are learned and executed by
the replicated state machines. These replicas typically play the roles
of acceptors (i.e., the processes that actually agree on a value) and
learners.  Paxos is resilience-optimum in the sense that it tolerates
the failure of up to $f$ acceptors from a total of $2f+1$
acceptors---to ensure progress---where a quorum of $f+1$ acceptors
must be non-faulty~\cite{Lam06b}.  In practice, replicated services
run multiple executions of the Paxos protocol to achieve consensus on
a sequence of values~\cite{chandra07}. An execution of Paxos is called
an instance.

An instance of Paxos proceeds in two phases. During the first phase, a
proposer that wants to submit a value selects a unique round number
and sends a prepare request to a group of acceptors (at least a
quorum).  Upon receiving a prepare request with a round number bigger
than any previously received round number, the acceptor responds to
the proposer promising that it will reject any future prepare requests
with smaller round numbers.  If the acceptor already accepted a
request for the current instance (explained next), it will return the
accepted value to the proposer, together with the round number
received when the request was accepted. When the proposer receives
answers from a quorum of acceptors, it proceeds to the second phase of
the protocol.

The communication pattern for Phase 2 of the protocol is
illustrated in Figure~\ref{fig:paxos}.  The proposer selects a value
according to the following rule.  If no acceptor in the quorum of
responses accepted a value, the proposer can select a new value for
the instance; however, if any of the acceptors returned a value in the
first phase, the proposer chooses the value with the highest round
number.  The proposer then sends an accept request with the round
number used in the first phase and the value chosen to at least a
quorum of acceptors.  When receiving such a request, the acceptors
acknowledge it by sending a message to the learners, unless the
acceptors have already acknowledged another request with a higher
round number.  Some implementations extend the protocol such that the
acceptor also sends an acknowledge to the proposer.  When a quorum of
acceptors accepts a value consensus is reached.

If multiple proposers simultaneously execute the procedure above for
the same instance, then no proposer may be able to execute the two
phases of the protocol and reach consensus.  To avoid scenarios in
which proposers compete indefinitely in the same instance, a
\emph{coordinator} process can be chosen.  In this case, proposers
submit values to the coordinator, which executes the first and second
phases of the protocol.  If the coordinator fails, another process
takes over its role.  Paxos ensures consistency despite concurrent
coordinators and progress in the presence of a single coordinator.

If the coordinator identity does not change between instances, then
the protocol can be optimized by pre-initializing acceptor state with
previously agreed upon instance and round numbers, avoiding the need
to send phase 1 messages~\cite{lamport98}. This is possible because
only the coordinator sends values in the second phase of the
protocol. With this optimization, consensus can reached in three
communication steps: the message from the proposer to the coordinator,
the accept request from the coordinator to the acceptors, and the
response to this request from the acceptors to the coordinator and
learners.

\begin{figure}[t]
\centering
   \begin{subfigure}[t]{.4\columnwidth}
     \centering
    \includegraphics[width=\columnwidth]{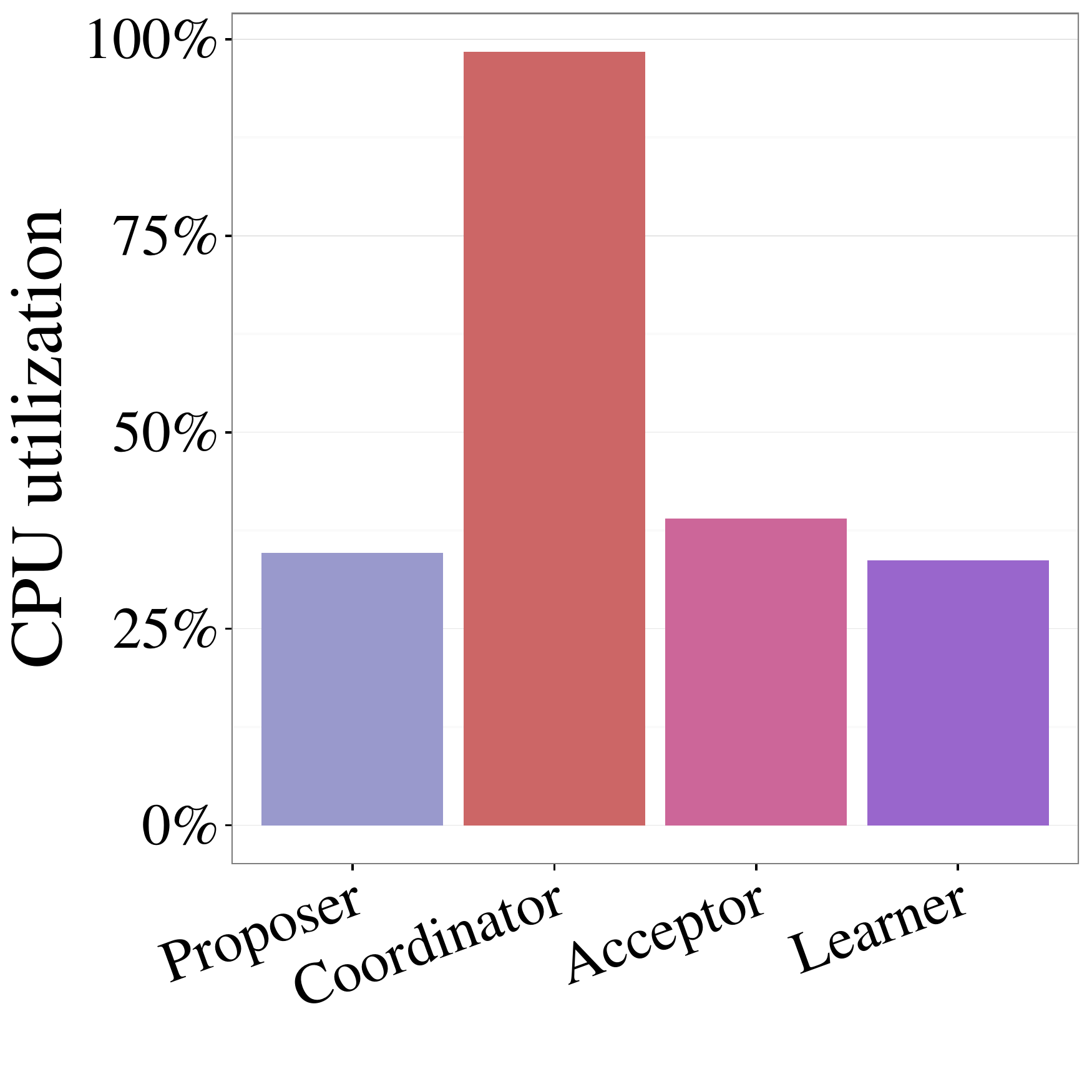}
    \caption{Coordinator bottleneck}
    \label{fig:bottleneck1}
   \end{subfigure}%
   \begin{subfigure}[t]{.4\columnwidth}
     \centering
      \includegraphics[width=\columnwidth]{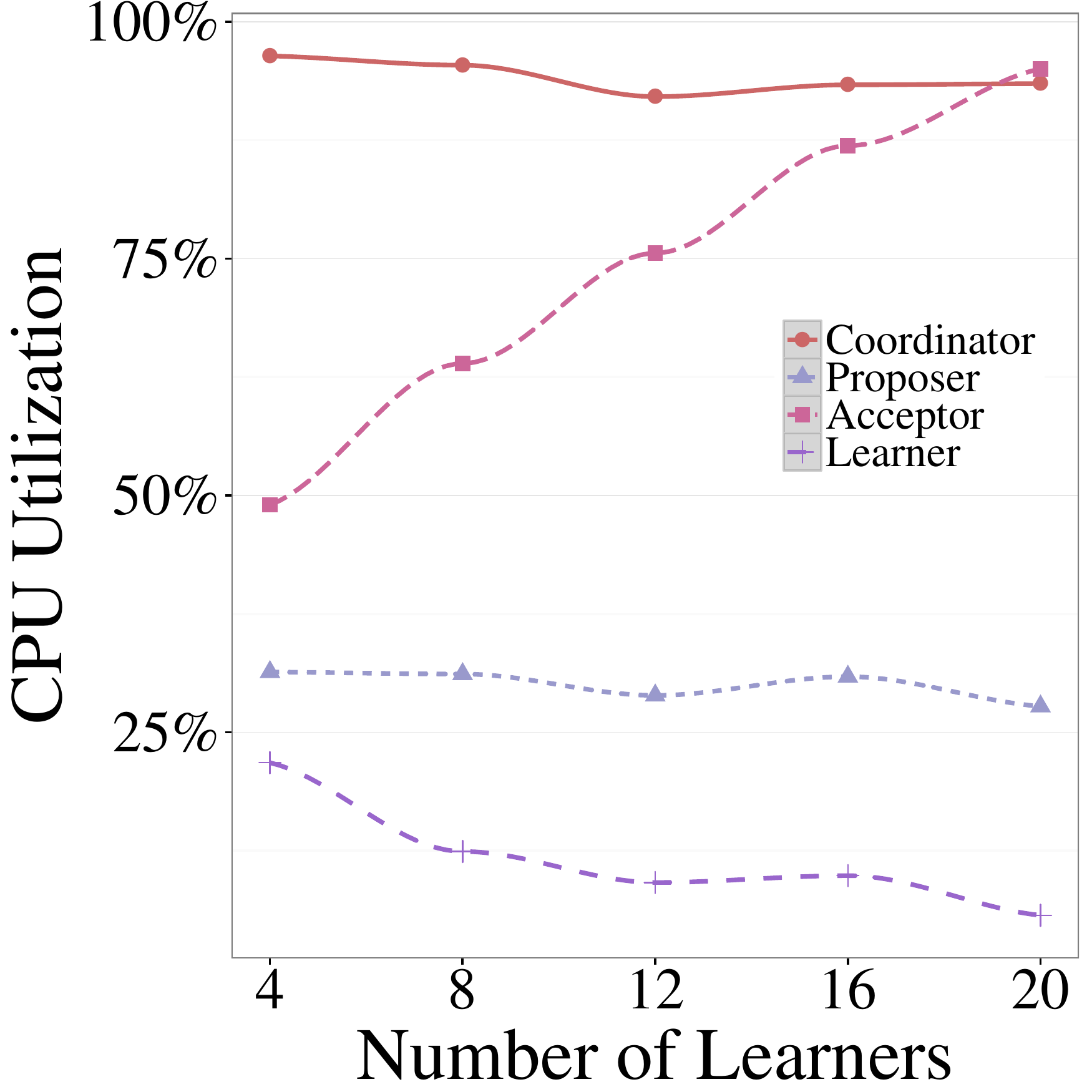}
        \caption{Acceptor bottleneck}
        \label{fig:bottleneck2}
   \end{subfigure}%
\caption{The coordinator and acceptor processes are the bottleneck in a software-based Paxos deployment. Acceptor utilization scales with the degree of replication.}
\end{figure}

\subsection{Motivation}


To investigate the performance bottleneck for a typical Paxos
 deployment, we measured the CPU utilization for each of the Paxos
 participants when transmitting messages at peak throughput.  As a
 representative implementation of Paxos, we used the open-source
\texttt{libpaxos}
library~\cite{libpaxos}. There are, of course, many Paxos
implementations, so it is difficult to make generalizations about
their collective behavior. However,
\texttt{libpaxos} is a faithful implementation of Paxos that
distinguishes all the Paxos roles. It has been extensively tested and
is often used as a reference Paxos implementation (e.g.,
\cite{istvan16,marandi14,Poke:2015,sciascia13}). Moreover,
\texttt{libpaxos} performs better than all the other available Paxos
libraries we are aware of under similar conditions~\cite{moraru13}.

In the initial configuration, there are seven processes
spread across three machines running on separate cores: one proposer that
generates load,
one proposer that is designated the coordinator, 
three acceptors, and two learners. The processes were distributed as follows:
\begin{itemize}
\itemsep -0.5em 
\item \emph{Server 1:} 1 proposer, 1 acceptor, 1 learner
\item \emph{Server 2:} 1 coordinator, 1 acceptor
\item \emph{Server 3:} 1 acceptor, 1 learner
\end{itemize}

\noindent
The details of the hardware are explained in
Section~\ref{sec:evaluation}. For brevity, we do not repeat them here.

The client application sends 64-byte messages to the proposer at a
peak throughput rate of 69,185 values/sec. The results, which show the
average utilization per role, are plotted in
Figure~\ref{fig:bottleneck1}. They show that the coordinator is
the bottleneck, as it becomes CPU bound.

We then extended the experiment to measure the CPU utilization for
each Paxos role as we increased the degree of replication by adding
additional learners.  The learners were assigned to one of three
servers in round-robin fashion, such that multiple learners ran on 
each machine.

The results, plotted in Figure~\ref{fig:bottleneck2}, show that as we
increase the degree of replication, the CPU utilization for acceptors
increases. In contrast, the utilization of the learners
decreases. This is because as the number of learners increases, the
throughput of the acceptor decreases, resulting in fewer messages sent
to the learners, and lower ultimately, lower learner utilization.

Overall, these experiments clearly show that the
coordinator and acceptor are performance bottlenecks for Paxos.  To
address these bottlenecks, CAANS moves coordinator and acceptor logic
into network hardware. Intuitively, this involves: (i) mapping Paxos
messages into a format that can be processed by network devices, and
(ii) modifying network devices to execute consensus logic.  The
next section presents the details of the CAANS design.

\section{Design}
\label{sec:design}

At a high-level, CAANS is similar to any other Paxos implementation,
in that there are four roles that participants in the protocol play:
proposers, coordinators, acceptors, and learners.

An instance of consensus is initiated when one of the proposers sends
a message to the coordinator. The protocol then follows the
communication pattern shown in
Figure~\ref{fig:paxos}. While
executing an instance of consensus, the various roles provide the
following functionality:

\noindent \textbf{Proposer.} Proposers provide an API to the application to
\emph{submit} a value, and initiate an instance of consensus.

\noindent \textbf{Coordinator.} A coordinator brokers requests on behalf of
proposers. The coordinator ensures that proposers will not compete for
a particular instance (thus ensuring that every instance terminates),
and imposes an ordering of messages.  The coordinator implements a
monotonically increasing sequence number, binding messages that come
from proposers to consensus instances.

\noindent \textbf{Acceptor.} Acceptors are responsible for choosing a single value
for a particular instance. For every instance of consensus, each
acceptor must ``vote'' for a value. The value can later be delivered
if a quorum of acceptors vote the same way.  Acceptors represent the
``memory'' of the protocol and must maintain the history of proposals
for which they have voted. This history ensures that acceptors never
vote for different values for a particular instance, and allows the
protocol to tolerate lost or duplicate messages.

\noindent \textbf{Learner.} Learners are responsible for replicating a value for a
given consensus instance. Learners receive votes from the acceptors,
and \emph{deliver} a value if a majority of votes are the same (i.e.,
there is a quorum).

\begin{figure}
\centering
 \includegraphics[width=0.4\columnwidth]{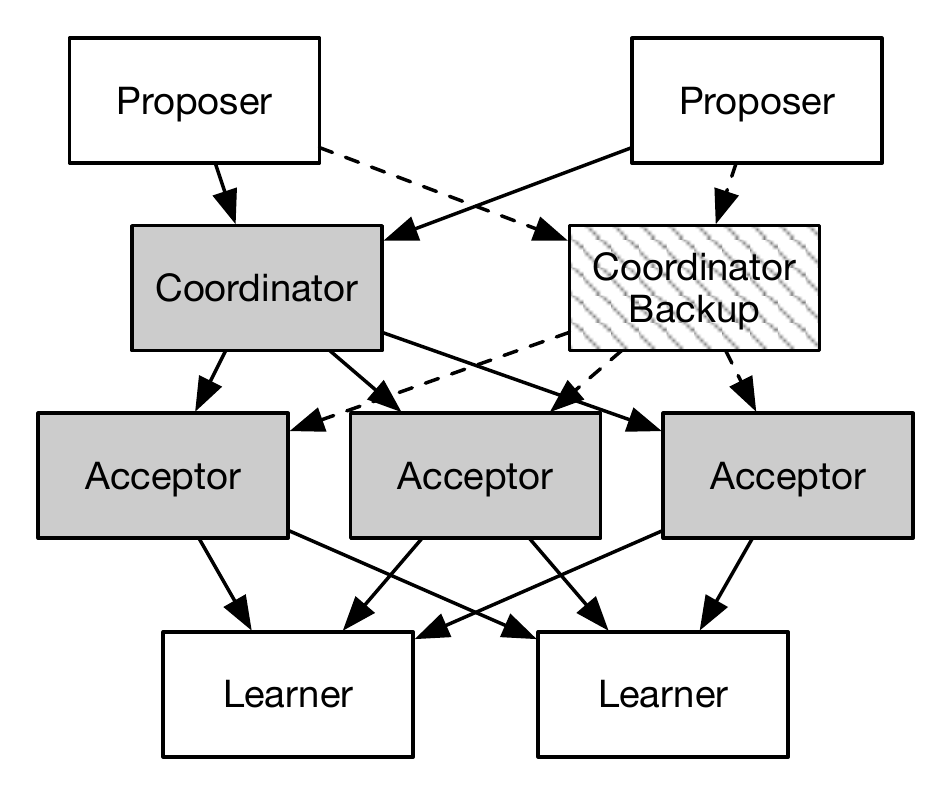}
\caption{A switch-based deployment of CAANS. Switch hardware is shaded grey, and commodity servers are colored white. The arrows represent communication. The coordinator backup may be deployed on either switch hardware of a commodity server. The dashed-arrows represent backup communication.}
\label{fig:netpaxos}
\end{figure}

\noindent
As per the discussion in Section~\ref{sec:background}, our design
assumes that a single coordinator is used by the proposers. Therefore,
CAANS acceptors can initialize their history of messages by implicitly
assuming an a priori execution of Phase 1, without the need for the
coordinator to actually execute Phase 1~\cite{gray06}. In other words,
acceptors initialize the round numbers in their history to the same
initial value used by the coordinator. This is a well-known
protocol optimization. We note that Paxos does not require that only
one participant act as coordinator at a time. As we will explain later
in details, if the coordinator fails, the system can elect a
new coordinator.

\subsection{Key Issues}
\label{sec:issues}

CAANS differs from typical Paxos deployments in that it moves some of the
consensus logic described above directly into the network. Intuitively, this
means (i) mapping Paxos messages into a format that can be processed by network
devices, and (ii) modifying network devices to execute consensus logic. However, 
implementing this design raises several key issues, including:

\begin{itemize}
\item What is the interface between the application and the Paxos implementation?
\item What logic should be implemented in hardware, and what logic should be implemented in software?
\item What is the format for Paxos messages?
\item How do memory constraints impact the protocol?
\item How should failures be handled?
\end{itemize}

\noindent
We discuss each of these topics in detail below.

\smartparagraph{Application interface.}
\label{sec:api}
One of the primary design goals for CAANS is to provide a drop-in
replacement for software-based implementations of
consensus. Therefore, the CAANS implementation of Paxos provides the
exact same API as a Paxos software library. The API consists of three
key functions, shown in Figure~\ref{fig:api}.

The \texttt{submit} function is called when the application using
Paxos sends a value. The application simply passes a character
buffer containing the value, and the buffer
size. The \texttt{paxos\_ctx} struct maintains Paxos-related state
across invocations (e.g., socket file descriptors).

The application must also register a callback function with the type
signature of \texttt{deliver}.  To register the function, the
application sets a function pointer in the
\texttt{paxos\_ctx} struct.
When a learner learns a value, it calls the
application-specific \texttt{deliver} function. The \texttt{deliver}
function returns a buffer containing the learned value, the size of
the buffer, and the instance number for the learned value.

The \texttt{recover} function is used by the application to discover a
previously agreed upon value for a particular instance of
consensus. The recover function results in the same sequence of Paxos
messages as the \texttt{submit} function. The difference in the API,
though, is that the application must pass the consensus instance
number as a parameter, as well as an application-specific no-op value.
The resulting \texttt{deliver} callback will either return the accepted
value, or the no-op value if no value had been previously accepted
for the particular instance number.

\begin{figure}[t]
\centerline{
\begin{lstlisting}
void submit(struct paxos_ctx * ctx, 
                char * value, 
                int size);

void (*deliver)(struct paxos_ctx* ctx, 
                     int instance, 
                     char * value, 
                     int size);

void recover(struct paxos_ctx * ctx,
	         int instance, 
                 char * value, 
                 int size);
\end{lstlisting}}
\caption{CAANS application level API.}
\label{fig:api}
\end{figure}  

\smartparagraph{Hardware/Software divide.}
\label{sec:hardware-software}
An important question for offering consensus as a network service
is: \emph{exactly what logic should be implemented in network
hardware, and what logic should be implemented in software?}

In the CAANS architecture, network hardware executes the logic
of \emph{coordinators} and \emph{acceptors}.  This choice allows
CAANS to address the bottlenecks identified in
Section~\ref{sec:background}.  Moreover, since the proposer and
learner code are implemented in software, the design facilitates the
simple application-level interface described above. The logic of each
of the roles is neatly encapsulated by communication boundaries.

Figure~\ref{fig:netpaxos} illustrates the CAANS architecture for a switch-based
deployment. In the figure, switch hardware is shaded grey, and commodity servers
are colored white. Note that a backup coordinator can execute on either a second
switch, or a commodity server, as we'll discuss below. We should also point out
that CAANS could be deployed on other devices, such as the programmable NICs
that we use in the evaluation.

\smartparagraph{Paxos header.}
\label{sec:header}
Network hardware is optimized to process packet headers. 
Since CAANS targets network hardware, it is a natural choice to map
Paxos messages into a Paxos-protocol header. The Paxos header follows
the transport protocol header (e.g., UDP), allowing CAANS messages to
co-exist with standard network hardware.

In a traditional Paxos implementation, each participant receives
messages of a particular type (e.g., Phase 1A, 2A), executes some
processing logic, and then synthesizes a new message that it sends to
the next participant in the protocol.

However, network hardware, in general, cannot craft new messages; they
can only modify fields in the header of the packet that they are
currently processing. Therefore, a network-based Paxos needs to map
participant logic into forwarding and header rewriting decisions
(e.g., the message from proposer to coordinator is transformed into a
message from coordinator to each acceptor by rewriting certain
fields). Because the message size cannot be changed at the switch,
each packet must contain the union of all fields in all Paxos
messages, which fortunately are still a small set.

\begin{figure}
\centerline{
\begin{lstlisting}[xleftmargin=5.0ex]
struct paxos_t {
  uint8_t msgtype;  
  uint8_t inst[INST_SIZE];
  uint8_t rnd;
  uint8_t vrnd;
  uint8_t swid[8]
  uint8_t value[VALUE_SIZE];   
};

\end{lstlisting}}
\caption{Paxos packet header.}
\label{fig:header}
\end{figure}  

Figure~\ref{fig:header} shows the CAANS packet header for Paxos
messages, written as a C struct. To keep the header small, the
semantics of some of the fields change depending on which participant
sends the message.  The fields are as follows: $(i)$ \texttt{msgtype}
distinguishes the various Paxos messages (e.g., phase 1A, 2A);
$(ii)$ \texttt{inst} is the consensus instance number; $(iii)$
\texttt{rnd} is either the round number computed by the proposer or the round
number for which the acceptor has cast a vote; \texttt{vrnd} is the
round number in which an acceptor has cast a vote;
$(iv)$ \texttt{swid} identifies the sender of the message; and
$(v)$ \texttt{value} contains the request from the proposer or the
value for which an acceptor has cast a vote.

A CAANS proposer differs from a standard Paxos proposer because before
forwarding messages to the coordinator, it must first encapsulate the
message in a Paxos header. Through standard sockets, the Paxos header
is then encapsulated inside a UDP datagram and we rely on the UDP
checksum to ensure data integrity.

\smartparagraph{Memory limitations.}
CAANS aims to support practical systems that use Paxos as a building
block to achieve fault tolerance. A prominent example of these are
services that rely on a replicated log to persistently record the
sequence of all consensus values. The Paxos algorithm does not specify
how to handle the ever-growing, replicated log that is stored at
acceptors. On any system, this can cause problems, as the log would
require unbounded disk space, and recovering replicas might need
unbounded recovery time to replay the log. The problem is potentially
more acute for CAANS, since the target hardware (e.g., switches and NICs)
has limited memory available. This imposes a practical concern:
\emph{how is the protocol impacted by memory constraints?}

A main issue to consider is what happens in case certain learners are
slow and have not participated in recent Paxos instances. A mechanism
is needed for these lagging learners to catch-up with leading
learners. Following the design of Google's Paxos~\cite{chandra07}, we
believe that an application using CAANS must checkpoint its state to
ensure that past consensus instances will not be needed.
As part of the checkpoint process, $f+1$ learners would need to inform
the acceptors that they can trim their log up to a particular instance
number. However, the
decisions of when and how to checkpoint are application-specific, and
we therefore do not include these as part of the CAANS implementation.
In ongoing work, we are exploring appropriate application-level
checkpoint mechanisms that are orthogonal to the work described in
this paper.

We note that there are a few CAANS-specific issues for applications
to consider. First, occasionally slow learners might need the ability
to find out what the value of a certain instance Paxos instance was.
This could happen for instance if a learner observes gaps in Paxos
instances and suspects that it is lagging behind but a fresh
checkpoint is not (yet) available. CAANS provides a \emph{recover} API
to applications that can be used to discover what value was accepted
for a given consensus instance. We cover this in more details when
discussing failure handling.

Second, the number of instances that can be tracked is bounded by the
memory available to CAANS acceptors to store the consensus history,
which is addressed by the \texttt{inst} field of the Paxos header.  On
the one hand, setting the field too big could potentially impact
performance since the coordinator and acceptor code must reserve
sufficient memory and make comparisons on this value. On the other
hand, if the \texttt{inst} field is too small, then the coordinator
and acceptors could run out of available instances in a short time.
For example, at a rate of 150,000 consensus instances executed per
second with 64-byte values, acceptors would fill a 1GB buffer in less
than two minutes. Reusing the acceptor's storage buffer means that
past instances in the buffer will no longer be available. Over time,
we expect that this limitation will become less of a concern, as the
amount of memory available on devices continues to increase.

\smartparagraph{Failure handling.}
\label{sec:failures}
The Paxos protocol does not provide guarantees against message loss. Again, the
details of how to cope with message loss depend on the application and the
deployment. We therefore do not include this functionality in CAANS.

In general, to cope with message loss, an application must provide a way to (i)
identify that a message was lost, and (ii) re-transmit the message. One common
approach is to use timeouts.  A
proposer that is also a learner should deliver its submitted value.
If the proposer times out before delivering the value, it assumes that
a message was lost (e.g., its request to the coordinator, the
coordinator's prepare message to the acceptors) and resubmits the
value.  This mechanism requires learners to detect and discard
duplicated delivered values, a situation that may happen if proposers
use timeouts that are too aggressive.

Learners that miss the decision of a consensus instance can find out
the value decided by contacting the acceptors through the
\emph{recover} API.  To find out the value decided in some instance
$k$, this API performs a new execution of the Paxos protocol (phase
1+2) while proposing a no-op value in consensus $k$ and waits for the
result.  If a value was decided in $k$, the learner will receive the
value, assuming a quorum of acceptors; otherwise, no-op will be decided in $k$ and discarded by the
application.

A faulty switch coordinator can be replaced by either another switch
or a coordinator in software running on a server that temporarily
assumes the role. If a software coordinator is chosen, then it can be
co-located with any of the proposers. The most important information
the new coordinator needs is the last consensus instance used.  This
information does not need to be accurate though.  If the next instance
the new coordinator uses is smaller than the last instance used by the
failed switch coordinator, then new values will not be accepted until
the new coordinator catches up with the last instance used.  If the
next instance used is greater than the last instance used, the
learners will see gaps in the sequence of learned values and will fill
in these gaps using the recover procedure described above.

Finally, we note that Paxos usually requires that acceptor storage is
persistent. In other words, if the acceptor fails, and then restarts,
it should be able to recover its durable state. Like prior
work~\cite{istvan16}, we assume that memory on network devices could
be made persistent through a variety of techniques, including using
battery-backed memory or newer NVRAM technology.

\subsection{Discussion}
The Paxos protocol does not guarantee or impose an ordering
on consensus instances. Rather, it guarantees that for a given
instance, a majority of participants agree on a value. So, for
example, the $i$-th instance of consensus need not complete before the
$(i+1)$-th instance.

Many applications may require stronger properties. For example, an
application might want to know if a given instance has not reached
consensus.  The process of detecting a missing instance and
re-initiating consensus depends on the details of the particular
application and deployment. For example, if proposers and learners are
co-located, then a proposer can observe if an instance has reached
consensus. If they are deployed on separate machines, then proposers
would have to employ some other process (e.g., using acknowledgments
and timeouts).

This observation begs the natural question:
\emph{why not implement functionality beyond consensus in the network?}
For example, one could imagine implementing state-machine replication,
or like Istv\'{a}n et al.~\cite{istvan16}, implement a replicated
key-value store.

Certainly, this is a topic that will require further research. However,
we believe that CAANS occupies a sweet-spot amongst the issues raised
by the end-to-end principle~\cite{saltzer84}: consensus is widely-applicable
to a large class of applications; the implementation provides significant
performance improvements; and it avoids implementing
application-specific logic in the network.

\section{Implementation}
\label{sec:implementation}

\begin{figure}[t]
\centering
 \includegraphics[width=0.4\columnwidth]{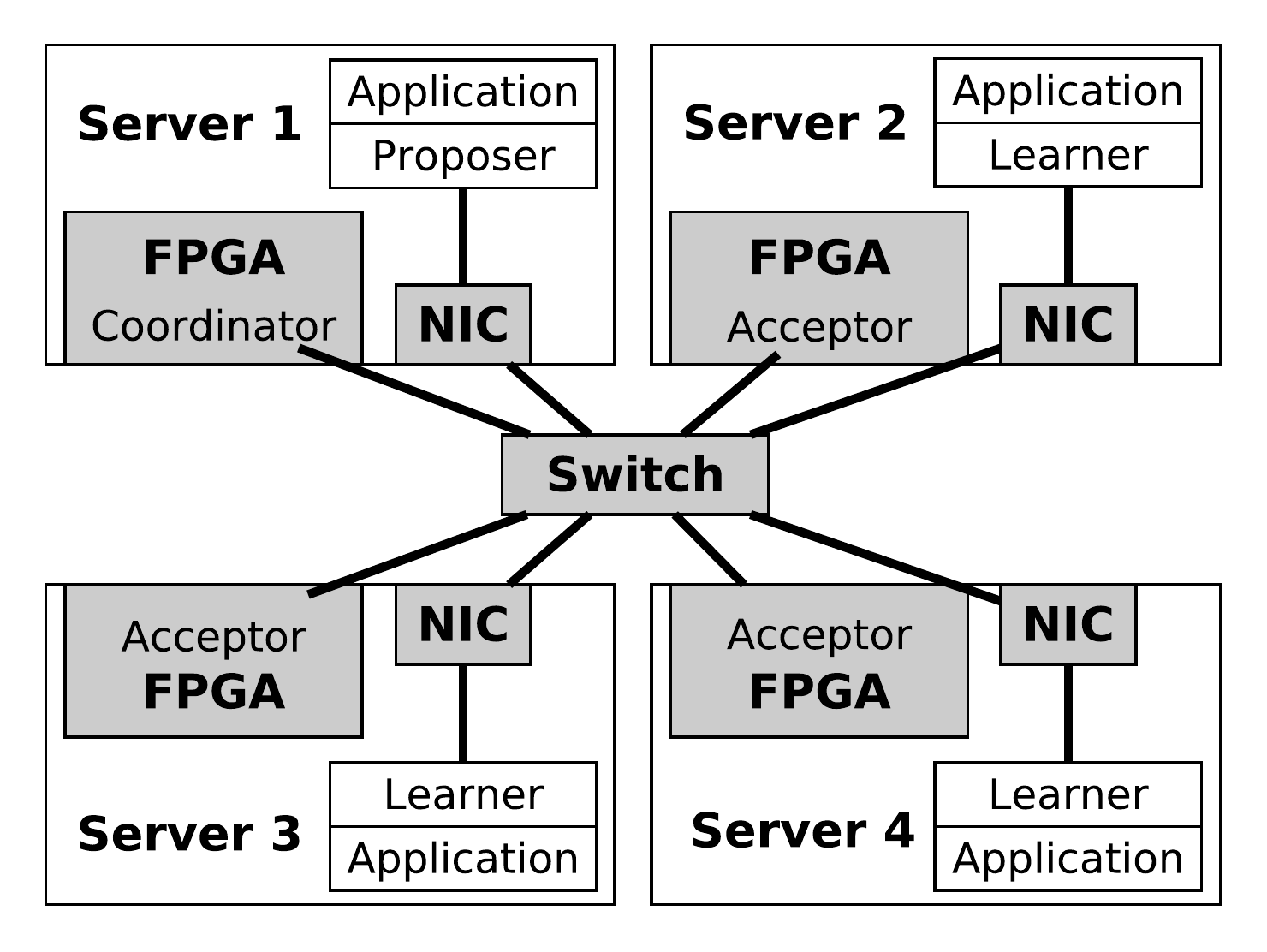}
\caption{An implementation of CAANS using FPGAs.}
\label{fig:fpgapaxos}
\end{figure}

We have implemented a prototype of CAANS. The proposer and learner
code are written in C, and the coordinator and
acceptor are written in P4~\cite{bosshart14}. All code, including
a demo that runs in a simulated environment, is publicly 
available with an open-source license\footnote{https://github.com/usi-systems/p4paxos}.

P4 is a high-level, data-plane programming language that can be compiled
to several hardware backends including FPGAs and network processors. 
The P4 implementation of CAANS follows the design of Dang et al.~\cite{dang16}. 

We have experimented with four different P4 platforms:
P4FPGA~\cite{wang15}, Netronome's Open-NFP~\cite{open-nfp}, Xilinx's
SDNet~\cite{sdnet}, and P4.org's compiler~\cite{p4org} which targets a
simulated environment (i.e., Mininet~\cite{mininet} and P4 Behavioral
Model switch~\cite{bm}).  Since P4 is relatively new and rapidly
changing, our experiments and results push the
state-of-the-art of P4 environments.  Specifically, none of the
compilers could generate a complete P4 source-to-target
solution. Each of the compilers required some additional
implementation effort, which we describe below.

\smartparagraph{P4FPGA.}
P4FPGA~\cite{wang15} is an open-source framework that compiles and
runs high-level P4 programs to various FPGA targets such as Xilinx and
Altera FPGAs. P4FPGA first transforms P4 programs into a data-plane
pipeline expressed as a sequence of hardware modules written in
Bluespec Verilog~\cite{bluespec}. P4FPGA then compiles these Bluespec
modules to FPGA firmware.  The acceptor and coordinator pipelines
include a parser, match table and/or action block, then deparser.
Since the P4FPGA compiler is a work-in-progress and is not fully
automated yet, we modified P4FPGA to implement any missing
functionality and optimized it with hand-written Bluespec code.  For
example, P4FPGA generated six match/action stages for the acceptor
pipeline, and three of those stages were not necessary, since the
match operations were hit on every incoming packet.

An FPGA has a limited number of on-chip block ram (BRAM) and
registers.  As a result, the CAANS implementation is constrained by
the amount of BRAMs available, which is the primary resource for
logging consensus values.  To scale to the large number of consensus
instances required for CAANS, we could leverage off-chip SRAM and
DRAM.  Of course, accessing off-chip memory in general leads to longer
processing delays.  Our current implementation uses on-chip memory
only.

\smartparagraph{Open-NFP.}
The Open-NFP~\cite{open-nfp} organization provides a set of tools for
developing network function processing logic in server networking
hardware. These tools include a P4 compiler that targets 10, 40 and
100GbE Intelligent Server Adapters (ISAs) manufactured by Netronome.

The Open-NFP compiler currently does not support register related
operations and cannot parse header fields larger than 32
bits. Moreover, while P4 version 1.0.2 allows users to define actions
for handling packets, the Open-NFP compiler supports a custom P4
syntax that declares actions with the \texttt{primitive\_action}
keyword, and implements actions in MicroC code external to the P4
program. The MicroC primitive actions can be used to extract packet
headers and to read or write header fields. 
The Open-NFP primitive actions are similar to ``external actions'' in
the proposed P4 version 1.2 specification. 

Therefore, to target the Netronome hardware devices, our P4
implementation of Paxos needed to be modified to gather all actions
and register declarations in a MicroC sandbox. Because the coordinator
and acceptor code require different register sizes, we had to
explicitly assign memory locations for registers. Additionally, we
used MicroC code to access hardware counters to compute the operation
latency.

\smartparagraph{Xilinx SDNet.}
Xilinx SDNet~\cite{sdnet} is a development environment for data plane
programmability that allows for scalability across the range of Xilinx
FPGAs.  SDNet compiles programs from the high-level
PX~\cite{brebner14} language to a data plane implementation on a
Xilinx FPGA target, at selectable line rates from 1G to 100G without
program changes.

A Xilinx Labs prototype P4 compiler works by translating from P4 to
PX, and then using SDNet to map this PX to a target FPGA.  A future
version in progress will compile P4 directly, rather than
cross-translating.  The SDNet-generated hardware IP blocks can then be
imported into a standard Xilinx Vivado project, in which they can be
connected to other hardware IP blocks. The resulting system is then
synthesized using the Vivado toolchain.  For NetFPGA
SUME~\cite{netfpgasume}, we needed to write a Verilog wrapper around
the SDNet standard interfaces to match them up with the expected
interfaces in the NetFPGA SUME reference switch design. One challenge
in this development effort was synchronizing packet metadata with the
packets that flow through the data plane pipeline.

The Xilinx Labs P4 compiler is currently under active development,
and evolving towards the future P4 version 1.2 specification.  To use
the compiler, we needed to modify our P4 version 1.0 implementation
to adopt expected version 1.2 features. The biggest difference was the
use of ``external methods'' to replace stateful register operations.

\smartparagraph{P4.org and P4 Behavioral Model}
In addition to the compilers described above, we also have used the
P4.org compiler~\cite{p4org} which targets a simulated environment
(i.e., Mininet~\cite{mininet} and P4 Behavioral Model
switch~\cite{bm}). As a software demo, we deploy one coordinator and
three acceptors as switches running CAANS, with a proposer and learners
running as Mininet hosts. An HTTP server runs on top of the proposer,
and a simple key-value store is deployed on each learner. Users connect
to the system using a web-interface to put and get values to the
replicated key-value store. Users can experiment with failure scenarios
using mininet commands to add and remove hosts or connections.

\section{Case Study: Replicated LevelDB}

As an example application of CAANS, we have used it to create a
distributed storage system backed by LevelDB~\cite{leveldb}. LevelDB
is a popular open-source, on-disk, key-value store implemented as a
library, rather than a server. LevelDB is used by a number of
applications, including Google's Chrome to store client-side data, and
BitCoin to store blockchain metadata~\cite{bitcoin}. Our own
implementation is inspired by Riak~\cite{riak}, which uses LevelDB as
one of its storage back-ends and Paxos for consensus.

Our application consists of two components: a client and a server.
The client code links against the CAANS proposer library.  Users of
the client submit get, put, and delete operations, which are then
serialized, and submitted to the storage servers via calls
to \texttt{submit}. The proposer library crafts the CAANS message with
the Paxos packet header, and sends the message to the coordinator.

Multiple servers run in parallel, each with their own instance of
LevelDB.  The storage is replicated. The
server code links against the CAANS learner library. When the learner
delivers a value, it invokes the application-specific callback
function.  The function returns the accepted value, which contains the
serialized request for the LevelDB instance. The server side code is
responsible for de-serializing the request, and invoking the
appropriate method from the LevelDB API. CAANS ensures that all replicas
are consistent.

It is worth emphasizing that the client and server components need
only interact with the consensus service via the small API described
in Section~\ref{sec:design}. The CAANS API is the same as any other
software-based implementation of Paxos.  No code from LevelDB needed
to be modified.  Thus, users of CAANS need only link against a
different library.

\section{Evaluation}
\label{sec:evaluation}

\begin{figure*}[t]
   \begin{subfigure}[t]{.25\textwidth}
     \centering
     \includegraphics[width=\columnwidth]{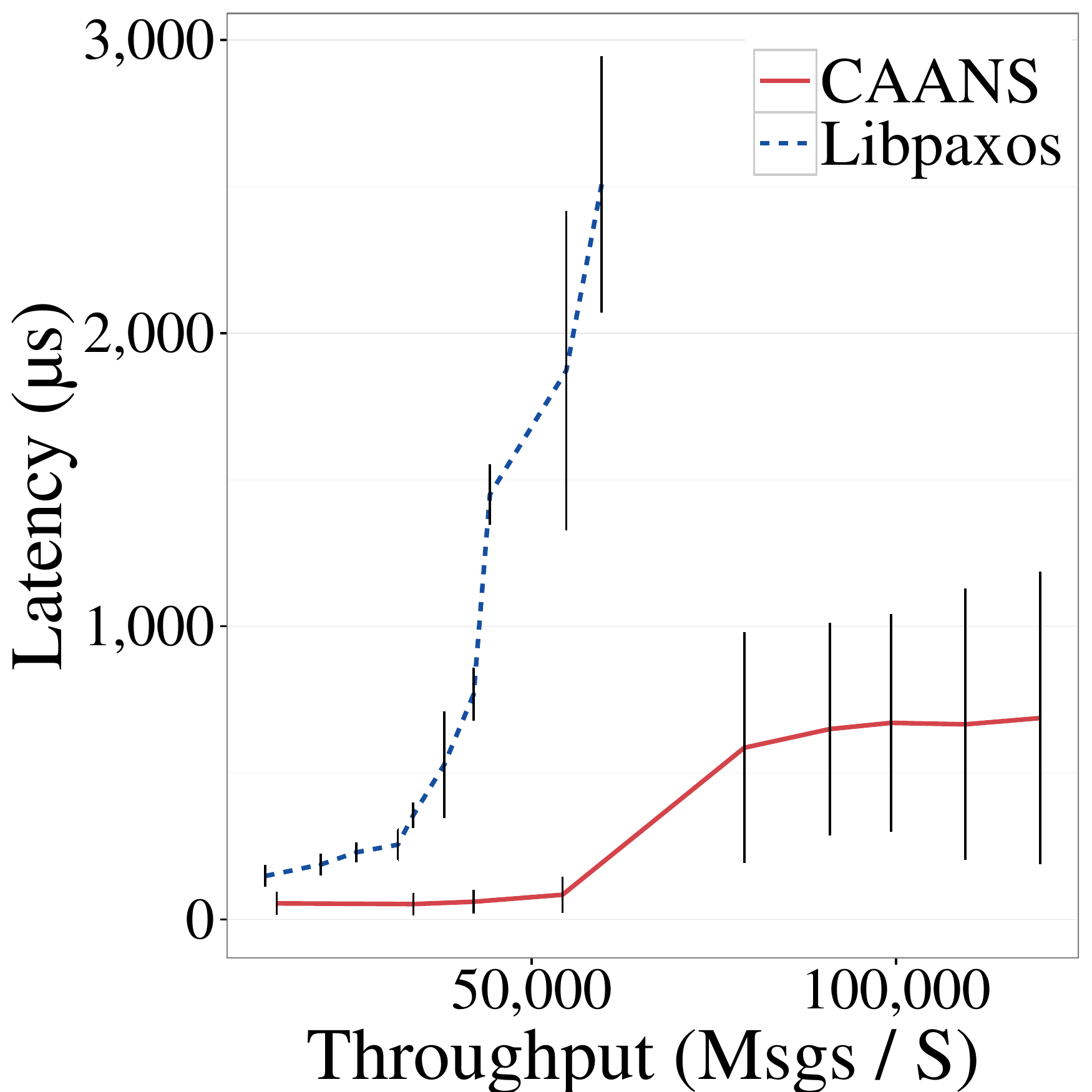}
     \caption{Throughput vs. latency}
     \label{fig:tput-vs-latency}
   \end{subfigure}%
   \begin{subfigure}[t]{.25\textwidth}
     \centering
      \includegraphics[width=\columnwidth]{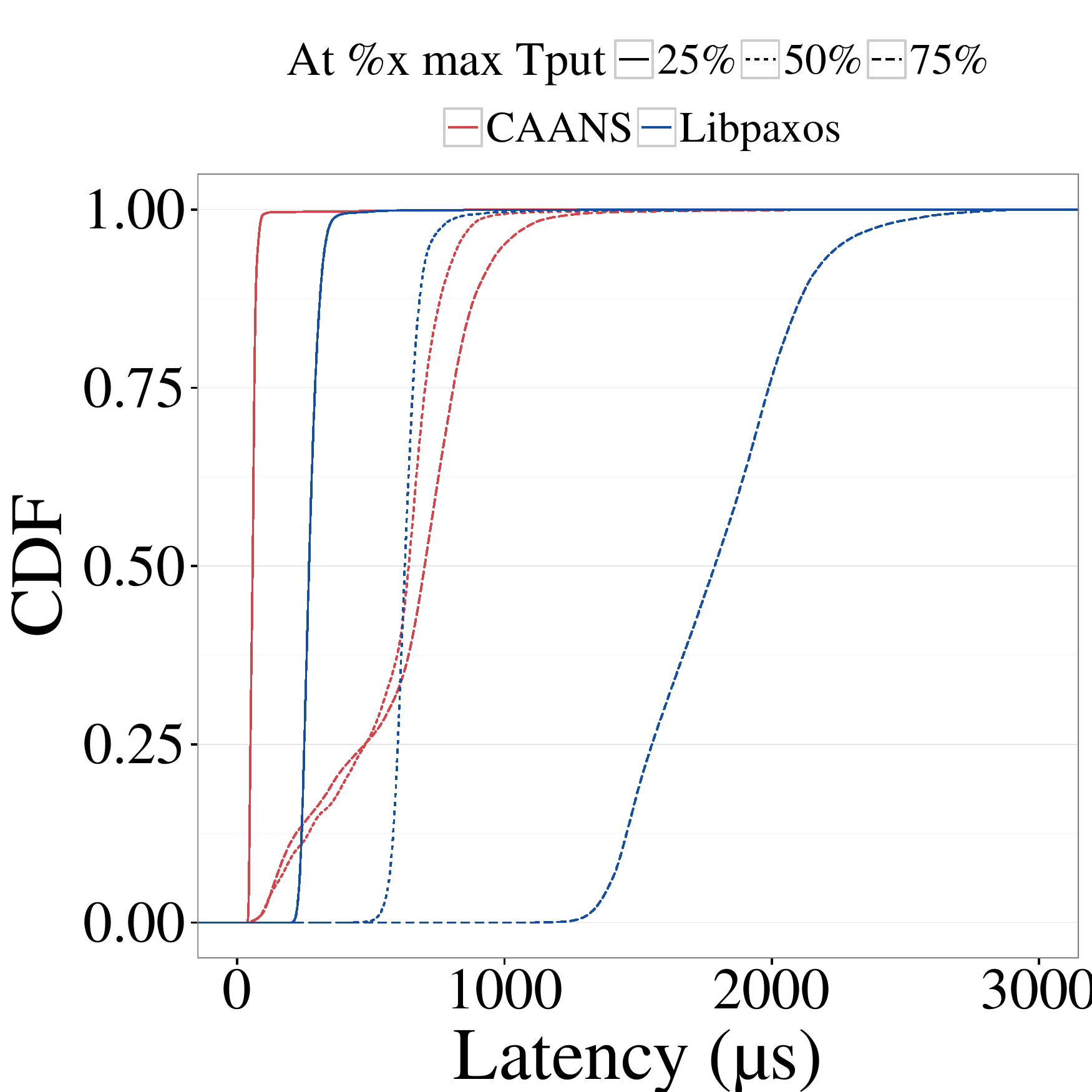}
      \caption{Latency CDF.}
      \label{fig:latency-cdf}
    \end{subfigure}%
   \begin{subfigure}[t]{.25\textwidth}
     \centering
      \includegraphics[width=\columnwidth]{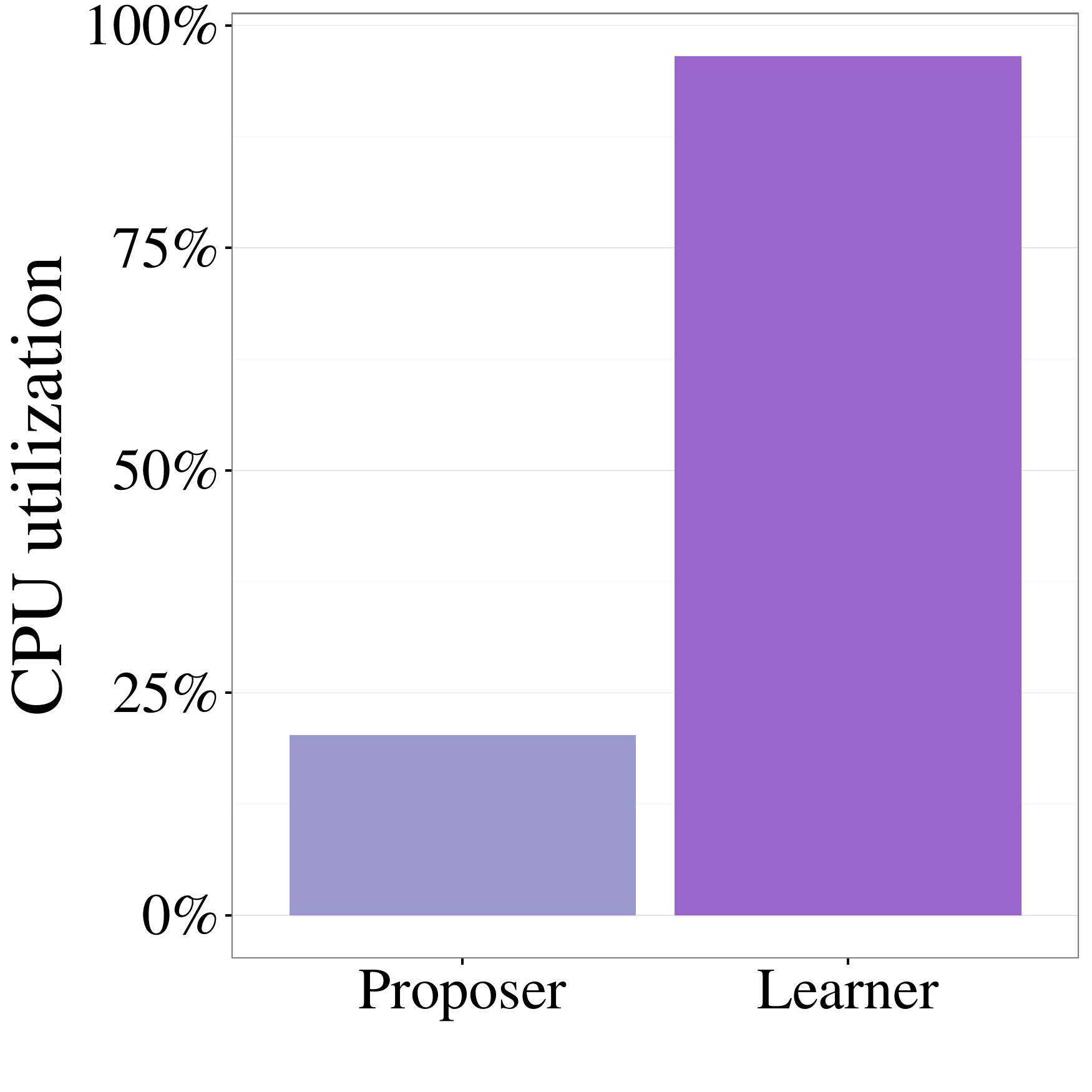}
      \caption{CPU Utilization.}
      \label{fig:util-caans}
      \end{subfigure}%
   \begin{subfigure}[t]{.25\textwidth}
     \centering
 \includegraphics[width=\columnwidth]{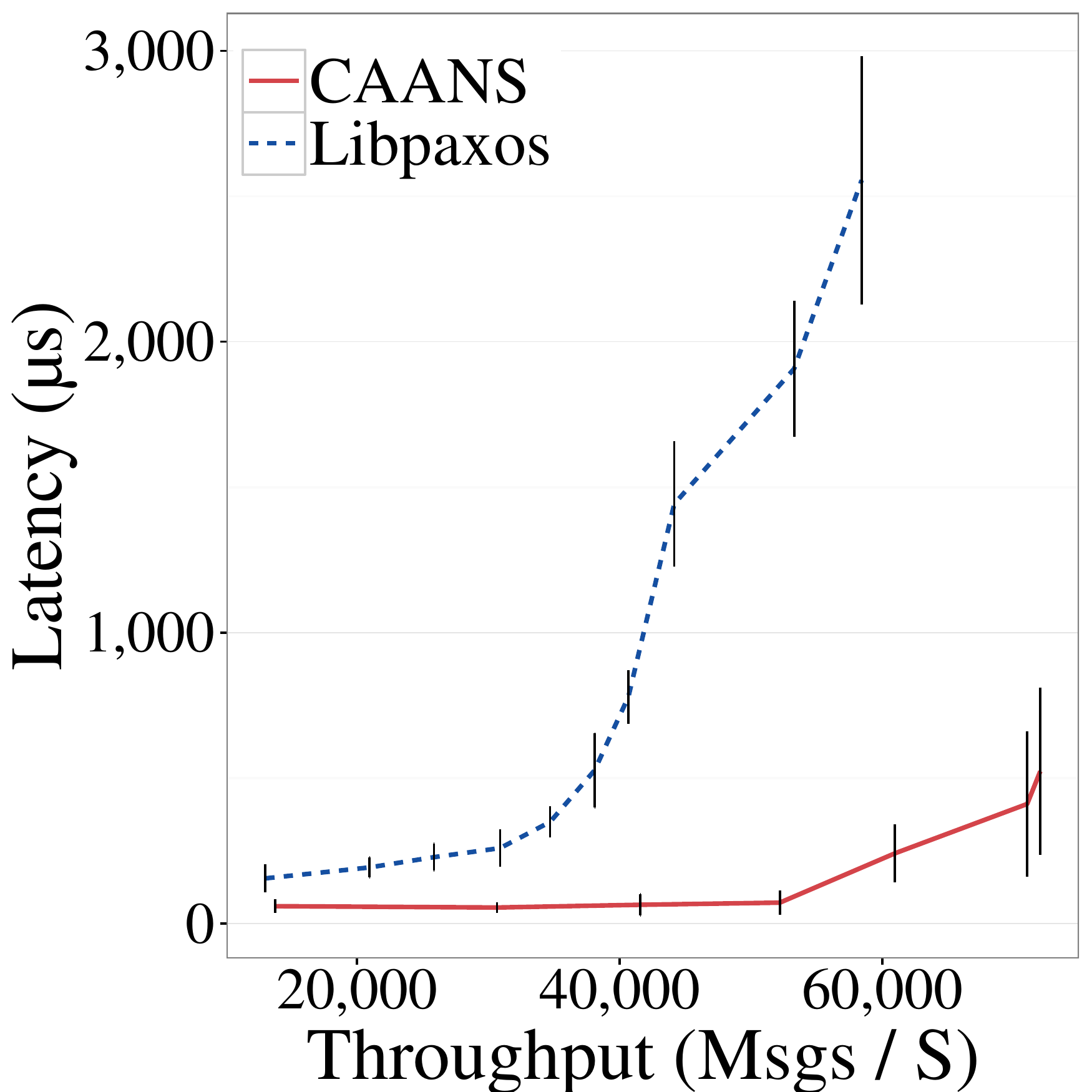}
 \caption{Throughput vs. latency (w/ LevelDB).}
 \label{fig:tput-vs-latency-leveldb}
\end{subfigure}%
\caption{The end-to-end performance of CAANS compared to libpaxos: (a) throughput vs. latency without, (b) latency CDF, (c) CAANS CPU Utilization, and (d) throughput vs. latency with LevelDB. CAANS demonstrates higher throughput, lower latency, and more predicable latency. Moreover, CAANS moves the processing bottleneck to the learner component, suggesting that even better performance should be possible.}
\end{figure*}

Our evaluation of CAANS explores three questions:
(i) What is the absolute performance of individual CAANS components?
(ii) What is the
end-to-end performance of CAANS as a system for providing consensus?
And, (iii) What is the performance under failure?

Overall, the results are promising. CAANS dramatically increases
throughput and reduces latency for end-to-end performance when
compared to software-only implementations.  Our experiments
show that CAANS shifts the processing bottleneck from coordinators and
acceptors to learners and the application itself. The absolute
performance of the hardware-based coordinators and acceptors suggests
that further performance gains would be possible with improvements in
learners and application design, pointing to future areas of research.
Moreover, we have evaluated CAANS on four different target devices
using three different compilers and toolchains\footnote{We don't use the P4.org compiler for performance measurements because it targets a simulated environment.}.  This demonstrates that
CAANS is portable across a diversity of network devices.

\subsection{Hardware Configuration}

We ran our end-to-end experiments on a testbed with four Supermicro
6018U-TRTP+ servers and a Pica8 P-3922 10G Ethernet switch, connected
in the topology shown in Figure~\ref{fig:fpgapaxos}.  The servers have
dual-socket Intel Xeon E5-2603 CPUs, with a total of 12 cores running
at 1.6GHz, 16GB of 1600MHz DDR4 memory and two Intel 82599 10 Gbps
NICs. We installed one NetFPGA SUME~\cite{netfpgasume} board in each
server through a PCIe x8 slot, though NetFGPAs behave as stand-alone
systems in our testbed. Two of the four SFP+ interfaces of the NetFPGA
SUME board and one of the four SFP+ interfaces provided by the 10G
NICs are connected to Pica8 switch with a total of 12 SFP+ copper
cables.  The servers were running Ubuntu 14.04 with Linux kernel
version 3.19.0.

In addition to the above hardware testbed, we also compiled CAANS
coordinator and acceptor P4 programs on the following target devices:
\begin{itemize}
\itemsep -0.5em 
\item A Netronome AgilIO-CX 1x40GbE Intelligent Server, with a  Netronome NFP-4000 processor, and 2GB of DDR3 DRAM memory
\item An Alpha Data ADM-PCIE-KU3: 2x40G NIC
\item A Xilinx VCU109: 4x100G line card
\end{itemize}





\subsection{Performance of CAANS components}

The first set of experiments explores the performance of CAANS
coordinator and acceptor logic. We focus on measuring throughput and
latency for a diverse set of hardware. Overall, the experiments show
that CAANS coordinators and acceptors can saturate a 10Gbps link, and
that the latency overhead for Paxos logic is little more than
forwarding delay.

\smartparagraph{Max achievable throughput.}
To measure the maximum rate at which coordinators and acceptors can
process consensus messages, we benchmarked their implementations on
the NetFPGA SUME using P4FPGA. We configured
a P4FPGA-based hardware packet generator and capturer to send 102
bytes\footnote{Ethernet header (14B), IP header (20B), UDP header
(8B), Paxos header (44B), and Paxos payload (16B)} consensus
messages to each of acceptor and coordinator, then captured and
timestamped each message measuring maximum receiving rate.

The results show that both acceptor and coordinator can process
over 9 million consensus messages per second, which is close to the
line-rate of 102 byte packets ($\approx$ 9.3M). We performed the same
benchmark with the Netronome card and measured the maximum throughput
at 2 million consensus messages per second. We suspect that
a higher throughput rate is possible with the Netronome card,
but that performance is limited by our
software license, which only provided evaluation access.


\begin{table}[t]
\centering
\begin{tabular}{ | l | c | c | c | c | }
\hline 
& P4FPGA & SDNet & Netronome \\ \hline
Forwarding & 0.37 $\mu$s & 0.73 $\mu$s & - \\ \hline
Acceptor & 0.79 $\mu$s & 1.44 $\mu$s & 0.81$\pm$0.01 $\mu$s \\ \hline
Coordinator & 0.72 $\mu$s & 1.21 $\mu$s  & 0.33$\pm$0.01 $\mu$s\\ \hline
\end{tabular}
\caption{Measured latency for acceptors and coordinators on NetFPGA SUME
with P4FPGA @ 250MHz,  NetFPGA SUME with SDNet @ 200MHz, and Netronome.}
\label{tab:hardware-latency}
\end{table}

\smartparagraph{Hardware latency.}
To quantify the processing overhead added by executing Paxos in
network hardware, we measured the pipeline latency of forwarding 
with and without Paxos. 
In particular, we computed the difference between two timestamps, one when
the first word of a consensus message entered the pipeline and the
other when the first word of the message left the pipeline.

Table~\ref{tab:hardware-latency} shows the results.  The first row
shows the results for forwarding without Paxos logic.  The latency was
measured between the parser and deparser stages of the forwarding
pipeline.  The results show that the latency for forwarding was 0.37
and 0.73 us for P4FPGA and SDNet, respectively.  We were not able to measure
the forwarding latency for Netronome, as we do not have access
to parser or deparser timestamps in Netronome.  The second and third
rows show the pipeline latency for an acceptor and a coordinator.
Acceptors of P4FPGA and SDNet represent a modest increase in latency of 
0.42 and 0.71 us over forwarding. 
The difference in latency between a coordinator and
acceptor was due to the complexity of operations of consensus
messages: acceptors do more work per message, hence the higher
latency. Lastly, we observed that the latency of the FPGA
based targets were constant due to their pipelines being a constant
number of stages; whereas the NPU-based Netronome target had some
measurable variance, which is due to the internal architecture and
scheduling of threads in NPU.

\smartparagraph{Computed throughput and latency.}
In addition to the above measured latencies, we also computed
throughput and latency for three different target devices when
compiling with SDNet.  SDNet can report precisely how many clock
cycles it takes a packet to go through from input to output.  One
can then just multiply this by the clock rate of the target board, and
have reliable latency numbers.  For throughput, one can take the
chosen bus width and the clock rate, and adjust for end-of-packet
fragmentation.  The computed throughputs and latencies are shown in
Table~\ref{tab:xilinx}. The latency differences here result
predominantly from increasing clock rates on the three platforms, 
and are not due to significant differences in pipeline length.


\begin{table*}
\footnotesize
\begin{center}
\begin{tabular}{|c|c|c|c|c| }
  \hline                                                  
  Device & Coordinator latency & Acceptor latency & Throughput & Clock rate\\
  \hline
  Digilent NetFPGA SUME: 4x10G port switch & 1.280 $\mu$s & 1.515 $\mu$s & 60M packets/s & 200MHz \\
  Alpha Data ADM-PCIE-KU3: 2x40G NIC & 1.024 $\mu$s & 1.212 $\mu$s &  60M packets/s & 250MHz \\
  Xilinx VCU109: 4x100G line card & 0.867 $\mu$s & 1.023 $\mu$s  & 150M packets/s & 300MHz \\
    \hline                                                  
\end{tabular}
\end{center}
\caption{Computed latencies and throughput for code generated from Xilinx SDNet.}
\label{tab:xilinx}
\end{table*}

\begin{figure}[t]
\centering
   \begin{subfigure}[t]{.35\columnwidth}
     \centering
    \includegraphics[width=\columnwidth]{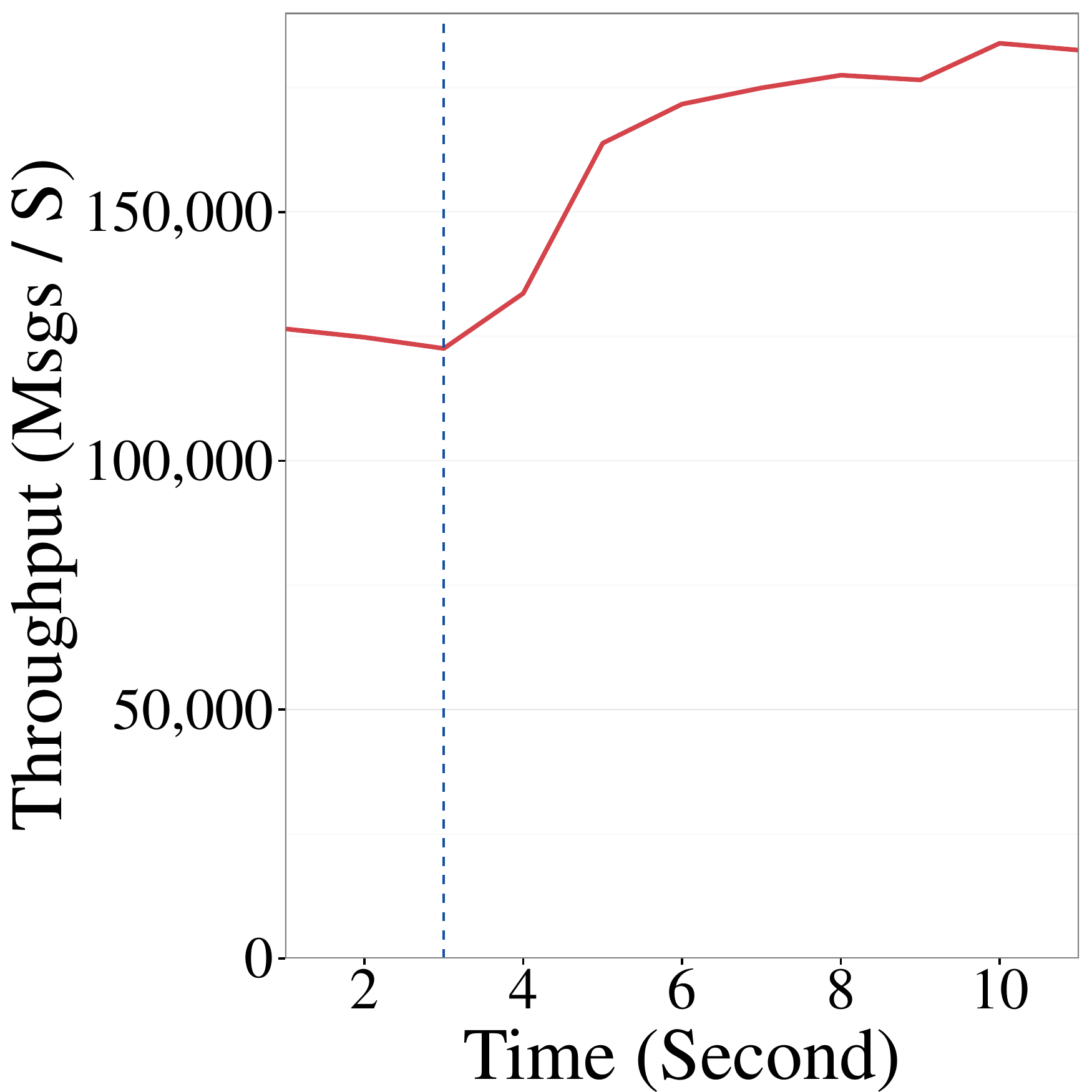}
    \caption{Acceptor failure}
    \label{fig:acceptor-failure}
   \end{subfigure}%
   \begin{subfigure}[t]{.35\columnwidth}
     \centering
      \includegraphics[width=\columnwidth]{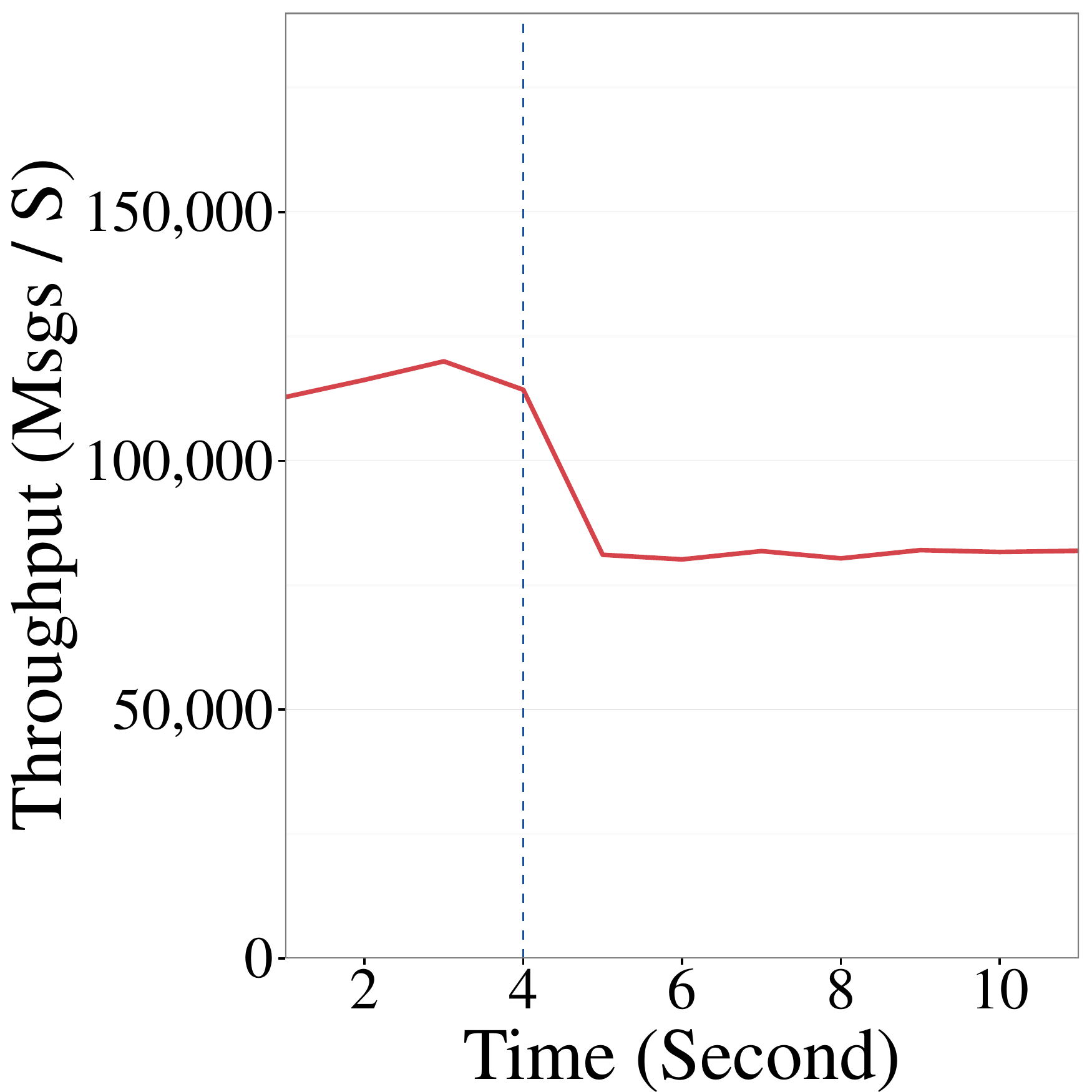}
        \caption{Coordinator failure}
        \label{fig:coordinator-failure}
   \end{subfigure}%
\caption{The performance of CAANS when (a) an acceptor fails, and (b) when the hardware coordinator fails and is replaced by a software coordinator.}
\label{fig:failure}
\end{figure}

\smartparagraph{Resource utilization.}
To evaluate the cost of implementing Paxos logic on FPGAs, we report
resource utilization on NetFPGA SUME using P4FPGA. An FPGA contains a
large number of programmable logic blocks: look-up tables (LUTs),
registers and block rams (BRAMs). In NetFPGA SUME, we implemented P4
stateful memory with on-chip BRAM to store the consensus history. As
shown in Table~\ref{tab:resource}, current implementation uses 54\% of
available BRAMs, out of which 35\% are used for stateful memory.  As
mentioned in Section~\ref{sec:implementation}, we could scale up the
current implementation in NetFPGA SUME by using large, off-chip DRAM
at a cost of higher memory access latency. It is possible to hide DRAM
access latency without significant affect overall
throughput~\cite{istvan16}. We note that cost may differ with other
compilers and targets.

\begin{table}[t]
\centering
\begin{tabular}{ | l | r | }
\hline
  LUTs & 84674 / 433200 (19.5\%) \\ \hline
  Registers & 103921 / 866400 (11.9\%) \\ \hline
  BRAMs & 801 / 1470 (54.4\%) \\ \hline
\end{tabular}
\caption{Resource utilization on NetFPGA SUME with P4FPGA with 65,535 Paxos instance numbers.}
\label{tab:resource}
\end{table}

\subsection{End-to-End Performance}
\label{sec:end-to-end}

To evaluate the end-to-end performance, we compare the throughput and
latency for transmitting consensus messages when using CAANS and the
software-based \texttt{libpaxos} mentioned in
Section~\ref{sec:background}.

Figure~\ref{fig:fpgapaxos} shows the deployment for CAANS. As an
application, we used a simple echo server. Server 1 runs a
multi-threaded client process and a single proposer process. Servers
2, 3, and 4 run learner processes and the echo server. The deployment
for \texttt{libpaxos} is similar, except that the coordinator and
acceptor processes run in software on their servers, instead of
running on the FPGA boards.

Each client thread submits a message with the current timestamp
written in the value. When the value is delivered by the learner, a
server program retrieves the message via a deliver callback function,
and then returns the message back to the client.  When the client gets
a response, it immediately submits another message. The latency is
measured at the client as the round-trip time for each
message. Throughput is measured at the learner as the number of
deliver invocations over time. 

To push the system towards higher message throughputs, we increased
the number of threads running in parallel at the client. The number
of threads, $N$, ranged from 1 to 22 by increments of 1. We stopped
measuring at 22 threads because the CPU utilization on the learners
reached 100\%. For each value of $N$, the client sent a total of 2
million messages (i.e., unique instances). We repeat this for three
runs, and report the mean latency and throughput over the three runs.

\smartparagraph{Throughput and latency.}
The results are shown in Figure~\ref{fig:tput-vs-latency}. The
experiment shows that CAANS results in dramatic improvements in
throughput and latency.  While \texttt{libpaxos} is only able to
achieve a maximum throughput of 59,604 messages per second, CAANS
reaches 134,094 messages per second, a 2.24x improvement.  The lowest
latency for \texttt{libpaxos} occurs at the lowest throughput
rate, and is 157$\mu$s. However, the latency increases significantly
as the throughput increases, reaching 2,510$\mu$s. In contrast, the
latency for CAANS starts at only 52$\mu$s, and is never higher than
686$\mu$s.

\smartparagraph{Predictability.}
CAANS messages also demonstrate more predictability than
\texttt{libpaxos}. Since applications typically do not run at maximum
throughput, we report the results for when the application is sending
traffic at 25, 50, and 75\% of the maximum throughput rates. Note that
the maximum throughput rates for CAANS and \texttt{libpaxos} are very
different. Figure~\ref{fig:latency-cdf} shows the latency
distribution. Table~\ref{tab:latency-dev} shows the average and
standard deviation for latencies for both CAANS and \texttt{libpaxos}.
We see that the latency measurements are much more stable for CAANS.

\begin{table}[t]
\centering
\begin{tabular}{| c | c | c | c | }
 \hline
    & \% max Tput & avg latency ($\mu$s) & std latency ($\mu$s) \\   
\hline
   CAANS  & 25\%  & 60.8     & 38.4   \\
   CAANS  & 50\%  & 582.5    & 389.2   \\
   CAANS  & 75\%  & 653.9    & 385.1  \\
\hline
libpaxos  & 25\%  & 275.9    & 92.7   \\
libpaxos  & 50\%  & 637.4    & 145.4  \\
libpaxos  & 75\%  & 1800.1   & 465.0  \\
\hline
\end{tabular}
\caption{The mean latency and standard deviation for CAANS and libpaxos.}
\label{tab:latency-dev}
\end{table}

\smartparagraph{Utilization.}
In Section~\ref{sec:background}, we showed that in a typical software
deployment, the coordinator and acceptors become bottlenecks for
consensus. We repeated the experiment with CAANS, measuring the CPU
Utilization at the proposer and learners when sending at maximum
throughput. The results are shown in Figure~\ref{fig:util-caans}. With
CAANS, we measured close to 100\% utilization on the learner.  This
explains the limit on maximum throughput for CAANS, as the performance
bottleneck is moved to the learner.

\smartparagraph{Application overhead.}
As a final end-to-end performance experiment, we measured the
throughput and latency for consensus messages for our replicated
LevelDB example application. The LevelDB instances were deployed on
the three servers running the learners. We followed the same
methodology as described above, but rather than sending dummy values,
we sent an equal mix of get and put requests.

The results are shown in Figure~\ref{fig:tput-vs-latency-leveldb}. The
maximum achievable throughput for CAANS is reduced to 75,825 messages
per second. This shows that the replicated application itself adds significant
overhead, and in fact, becomes the bottleneck when run with CAANS. In contrast,
there is no throughput change for the \texttt{libpaxos} deployment, where
we measured a  maximum throughput of 58,427 messages per second. There is no
difference from the previous experiment, because with \texttt{libpaxos}, the
coordinator is the bottleneck.

\subsection{Performance Under Failure}

As a final set of experiments, we repeated the throughput and latency
measurements from Section~\ref{sec:end-to-end} under two different
failure scenarios. In the first, one of the three acceptors fails. In
the second, the hardware coordinator fails, and the coordinator is
temporarily replaced with a software coordinator. Note that in the
second scenario, acceptor logic remains in hardware.  In both the
graphs in Figure~\ref{fig:failure}, the blue, vertical line indicates
the failure point.

Figure~\ref{fig:acceptor-failure} shows that the throughput of CAANS
increases after the loss of one acceptor. This is expected, since the
learner is the bottleneck, and it processes fewer messages after the
failure.

To handle the loss of a coordinator, we re-route traffic to a software
coordinator.  Figure~\ref{fig:coordinator-failure} shows that CAANS is
resilient to the failure, although the software-coordinator adds
additional processing overhead.  We note that CAANS could use a backup
hardware-based coordinator.  Unfortunately, we did not have access to
an additional FPGA. However, we expect that performance would not
decrease with a hardware coordinator.

\section{Related Work}
\label{sec:related}



Dang et al.~\cite{dang15} proposed the idea of moving consensus logic
in to network devices. Their work suggested two possible approaches:
$(i)$ implementing Paxos in switches, and $(ii)$ using a modified
protocol, named \emph{NetPaxos}, which solves consensus without
switch-based computation by making assumptions about packet ordering.
Their Paxos Made Switch-y paper~\cite{dang16} makes the implementation
of a switch-based Paxos concrete, by presenting a detailed discussion
of the source code written in P4~\cite{bosshart14}. This paper builds
on their work, both by providing a complete system, and by presenting
a thorough of the performance benefits of network-based consensus.

The work of Istv\'{a}n et al.~\cite{istvan16} describes an
implementation of Zookeeper's atomic broadcast written in Verilog. As
mentioned in the introduction, CAANS differs from their work on
several key design points: (i) CAANS implements a complete Paxos
protocol, as opposed to the simplified ZAB protocol, (ii) CAANS
provides an general API exposed to software applications, while
Istv\'{a}n et al. require that applications be implemented inside of
the FPGA, and (iii) CAANS is portable across a range of target
devices.

\smartparagraph{Network support for applications.}
Several recent projects have investigated leveraging network
programmability for improved application performance, focusing
especially on large-scale, data processing systems, such as Hadoop.
For example, PANE~\cite{ferguson13}, EyeQ~\cite{jeyakumar13a}, and
Merlin~\cite{soule14} all use resource scheduling to improve the job
performance , while NetAgg~\cite{mai14} leverages
user-defined \emph{combiner} functions to reduce network
congestion. These projects have largely focused on improving
application performance through traffic management. In contrast, this
paper argues for moving application logic into network devices.

Speculative Paxos~\cite{ports2015} uses predictable network behavior
to improve the performance of Paxos.  It uses a combination of
techniques to eliminate packet reordering in a data center, including
IP multicast, fixed length network topologies, and a single
top-of-rack switch acting as a serializer.


\smartparagraph{Replication protocols.}
Research on replication protocols for high availability is quite mature.
Existing approaches for replication-transparent protocols, notably protocols
that implement some form of strong consistency (e.g., linearizability,
serializability) can be roughly divided into three classes \cite{CBPS10}: (a)
state-machine replication~\cite{lamport78,schneider90}, (b) primary-backup
replication~\cite{OL88}, and (c) deferred update replication~\cite{CBPS10}.

Despite the long history of research in replication protocols, there exist very
few examples of protocols that leverage network behavior to improve
performance. The one exception of which we are aware are systems that exploit
\emph{spontaneous message ordering}, \cite{lamport06,PS02b,PSUC02}.  The idea is
to check whether messages reach their destination in order, instead of assuming
that order must be always constructed by the protocol and incurring additional
message steps to achieve it. This paper differs in that it implements a standard
Paxos protocol that does not make ordering assumptions.

\section{Conclusion}
\label{sec:conclusion}

Performance has long been a concern for consensus protocols. Twenty
years ago, researchers observed that you might not use consensus
in-band for systems with high demand~\cite{friedman96}.  Since that
time, there have been many proposals made to optimize the performance
of consensus protocols
(e.g.,~\cite{moraru13,Benz:2014,Lamport04,PS99,lamport06,ports15,marandi10,reed08}).
But still, as recently as five years ago, Bolosky et
al.~\cite{bolosky11} wrote that ``Conventional wisdom holds that Paxos
is too expensive to use for high-volume, high-throughput,
data-intensive applications.''

The performance benefits provided by CAANS can have a dramatic impact
for local-area and data-center deployments. The CAANS approach means
that coordinators and acceptors can fully saturate 10G links, and the
latency for transmitting consensus traffic is little more than simply
forwarding.

With CAANS, the bottleneck for consensus becomes the learners and the
application itself. We believe that this changes the landscape for
future replication research. The focus of future work should be on
investigating how applications can handle data at such high rates.

There have been some initial efforts in this direction already.  For
example, some approaches explore how to leverage the parallelism of
multi-core machines to improve the performance of
replication~\cite{bezerra14,marandi2013rethinking,kotla2004high,kapritsos2012all,Cui:2015}.
Bolosky et al.~\cite{bolosky11} focuses on avoiding disk
latency. NetMap~\cite{rizzo12} investigate ways to avoid overhead for
reading packets. These techniques are complimentary to CAANS, and we
expect efforts like this will become increasingly important.

In summary, the advent of flexible hardware and expressive data\-plane
programming languages will have a profound impact on networks. One
possible use of this emerging technology is to move logic
traditionally associated with the application layer into the network
itself. In the case of Paxos, and similar consensus protocols, this
change could dramatically improve the performance of data center
infrastructure, and open the door for new research challenges in the
design of consensus protocols.

\paragraph*{Acknowledgements.}

We thank Gordon Brebner from Xilinx for providing us with access to
the SDNet research prototype, and for performing latency and
throughput measurements using SDNet.  We also thank Jici Gao, Mary
Pham, and Bapi Vinnakotam from Netronome for providing us with access
to their hardware testbed, and for support with their compiler.
Finally, we are grateful to Gianni Antichi for sharing his expertise
with the NetFPGA SUME board. This research is partially supported by
European Union's Horizon 2020 research and innovation programme
under the ENDEAVOUR project (agreement No. 644960) and the SSICLOPS
project (agreement No. 644866). It is also partially supported by DARPA
CSSG (D11AP00266), NSF (1422544, 1053757, 0424422, 1151268, 1047540),
and Cisco.


\bibliography{main}
\bibliographystyle{abbrv} 

\end{document}